\documentclass[sigconf]{acmart}
\AtBeginDocument{%
  }

\setcopyright{none}
\renewcommand\footnotetextcopyrightpermission[1]{}
\usepackage{algorithmic}
\usepackage{graphicx}
\usepackage{textcomp}
\usepackage{xcolor}
\usepackage{multirow}
\usepackage{xspace}
\usepackage{tabularx}
\usepackage{enumitem}
\newcommand{\HWName}{MASQ\xspace}
\begin{document}

%%
%% The "title" command has an optional parameter,
%% allowing the author to define a "short title" to be used in page headers.
\title{\HWName: Accelerating Masked Diffusion via Stage-Wise Multi-Precision Quantization}

%%
%% The "author" command and its associated commands are used to define
%% the authors and their affiliations.
%% Of note is the shared affiliation of the first two authors, and the
%% "authornote" and "authornotemark" commands
%% used to denote shared contribution to the research.

\author{Seeyeon Kim}
\affiliation{%
  \institution{KAIST}
  \city{Daejeon}
  \country{Republic of Korea}}
\email{seeyakim@kaist.ac.kr}

\author{Jaehun Lee}
\affiliation{%
  \institution{KAIST}
  \city{Daejeon}
  \country{Republic of Korea}}
\email{jaehunlee@kaist.ac.kr}

\author{Sungyeob Yoo}
\affiliation{%
  \institution{KAIST}
  \city{Daejeon}
  \country{Republic of Korea}}
\email{sungyeob.yoo@kaist.ac.kr}

\author{Joo-Young Kim}
\affiliation{%
  \institution{KAIST}
  \city{Daejeon}
  \country{Republic of Korea}}
\email{jooyoung1203@kaist.ac.kr}
% \authornote{Corresponding author.}

%%
%% By default, the full list of authors will be used in the page
%% headers. Often, this list is too long, and will overlap
%% other information printed in the page headers. This command allows
%% the author to define a more concise list
%% of authors' names for this purpose.
% \renewcommand{\shortauthors}{Kim et al.}

%%
%% The abstract is a short summary of the work to be presented in the
%% article.
\begin{abstract}

Masked diffusion enables region-specific image synthesis but suffers from computational redundancy, since the entire image is processed each timestep even though only the masked region requires generation.
To address this, we introduce \HWName, a hardware–software co-designed accelerator for masked diffusion.
Our approach performs stage-wise MXINT8/4/2 precision assignment that dynamically reflects spatial and semantic importance, complemented by timestep-aware scheduling and optimized non-matrix operations.
\HWName features a block-wise multi-precision compute engine and mask management unit, efficiently handling our approach.
It achieves up to 16.06$\times$ and 5.39$\times$ speedup and 4.18$\times$ and 4.93$\times$ energy-efficiency gain over A100 and Orin NX, respectively, while preserving quality.
\end{abstract}

%%
%% The code below is generated by the tool at http://dl.acm.org/ccs.cfm.
%% Please copy and paste the code instead of the example below.
%%


%%
%% Keywords. The author(s) should pick words that accurately describe
%% the work being presented. Separate the keywords with commas.
%% A "teaser" image appears between the author and affiliation
%% information and the body of the document, and typically spans the
%% page.
% \begin{teaserfigure}
%   \includegraphics[width=\textwidth]{sampleteaser}
%   \caption{Seattle Mariners at Spring Training, 2010.}
%   \Description{Enjoying the baseball game from the third-base
%   seats. Ichiro Suzuki preparing to bat.}
%   \label{fig:teaser}
% \end{teaserfigure}

% \received{20 February 2007}
% \received[revised]{12 March 2009}
% \received[accepted]{5 June 2009}

%%
%% This command processes the author and affiliation and title
%% information and builds the first part of the formatted document.
\maketitle

\section{Introduction}
\label{section1}
Masked diffusion~\cite{sdedit, dreaminpainter, smartbrush, stablediffusion, editbench} has become a widely adopted approach for controllable image synthesis.
It provides localized control, where a user-provided mask guides the model to selectively edit specific regions while largely preserving the surrounding content.
This enables intuitive and interactive workflows for tasks such as image inpainting and editing.
These strengths arise from the underlying diffusion models~\cite{ddpm, ddim, cascadeddiffusion, beatgan, glide, stablediffusion, imagen, sdxl, dalle}, which generate images through iterative denoising and have demonstrated improved stability and expressiveness over generative adversarial networks~\cite{gan, karras2019style, karras2020analyzing, brock2018large}.
Such diffusion models now serve as the foundation of recent state-of-the-art (SOTA) image-generation systems such as Stable Diffusion~\cite{stablediffusion} and DALLE-3~\cite{dalle}.
These generative strengths enable practical masked diffusion workflows that form a core capability of modern interactive image-generation systems.

\begin{figure}[t]
    \centering
    \includegraphics[width=\linewidth]{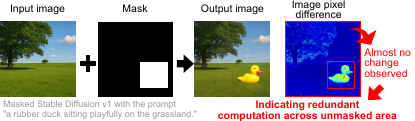}
    \vspace{-0.15in}
    \caption{Pixel-wise difference between input and output of masked diffusion}
    \label{fig:introduction}
    \Description{.}
    \vspace{-0.3in}
\end{figure}

Despite its ability to achieve region-specific and seamless synthesis, masked diffusion introduces a critical inefficiency.
As illustrated in Figure~\ref{fig:introduction}, a substantial portion of computation is spent on unmasked regions that remain mostly unchanged, as they are still processed uniformly in FP32 or FP16 formats.
To preserve global dependencies between masked and unmasked regions, existing methods recompute the entire feature tensor, ensuring global coherence and preventing visible seams.
SIGE~\cite{sige} attempts to mitigate this overhead by reusing intermediate representations of unmasked regions through a caching strategy.
% To alleviate this limitation, SIGE~\cite{sige} introduces a caching strategy that reuses intermediate representations of unmasked regions.
However, this approach relies on feature maps extracted from a pre-generated original image, incurring significant memory overhead and limiting general applicability.
Consequently, a more general and efficient solution is required to avoid unnecessary computation in masked diffusion.

To address the computational inefficiency of masked diffusion, we propose \HWName, a hardware–software co-designed accelerator that reduces redundant processing of unmasked regions. 
Unlike cache-based approaches, \HWName does not rely on pre-generated feature maps. 
Instead, it leverages the observation that unmasked regions change only minimally, which suggests that these regions can tolerate more aggressive low-precision computation.
Computing these regions at reduced precision, instead of skipping them entirely, preserves global coherence while substantially reducing computational cost.
Motivated by this, \HWName incorporates a mask-aware multi-precision strategy that reduces precision on unmasked regions while preserving high fidelity.

\begin{itemize}[leftmargin=*]
\item \textbf{On the algorithm level}, we propose a stage-wise multi-precision scheme that dynamically partitions the feature map into multiple stages based on the spatial and semantic importance derived from the mask.
It assigns higher precision MXINT8~\cite{mx} to masked or semantically critical regions while safely lowering precision to MXINT2 for less important areas, and further incorporates timestep-aware scheduling as well as precision-aware normalization and softmax.
This scheme effectively reduces redundant computation while preserving output fidelity.

\item \textbf{On the hardware level}, we design a specialized architecture featuring a flexible compute engine that supports multi-precision execution with block floating-point format MXINT8, MXINT4, and MXINT2.
Together with an efficient mask management unit that dynamically generates and updates multi-stage masks, this architecture enables \HWName to perform stage-wise masked execution and support a wide range of masked diffusion scenarios.

\item \HWName achieves up to 16.06$\times$ speedup and 4.18$\times$ improvement in energy efficiency over NVIDIA A100~\cite{nvidia_a100}, and up to 5.39$\times$ speedup and 4.93$\times$ efficiency gain on Jetson Orin NX~\cite{nvidia_nx}, all with negligible quality loss.
\end{itemize}
\section{Background and Motivation}
\label{section2}
\begin{figure}[t]
    \centering
    \includegraphics[width=0.95\linewidth]{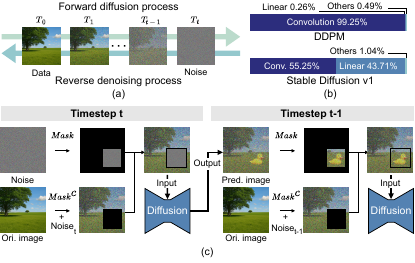}
    \vspace{-0.15in}
    \caption{(a) Overview of diffusion process (b) Computation breakdown by operation type in diffusion models (c) Overview of maksed diffusion process}
    \label{fig:diffusion_image_editing}
    \Description{.}
    \vspace{-0.25in}
\end{figure}
\subsection{Masked Diffusion}

Diffusion models have achieved remarkable success in image generation, synthesizing high-quality images starting from random noise.
At their core, they operate through an iterative reverse denoising process, driven by a U-Net~\cite{unet} that is applied repeatedly to gradually remove noise as illustrated in Figure~\ref{fig:diffusion_image_editing}(a).
The U-Net architecture consists of ResNet and attention blocks arranged across multiple resolutions, where feature maps are downsampled and then upsampled.
Each block is composed of matrix operations such as convolution and linear layers, along with non-matrix operations like normalization and softmax.
While the diffusion model achieves high-quality image generation through repeated denoising, this iterative process is computationally expensive, with matrix operations accounting for over 90\% of total operations in models like Stable Diffusion~\cite{stablediffusion}, as shown in Figure~\ref{fig:diffusion_image_editing}(b).

Masked diffusion, a key paradigm for tasks like image inpainting and editing~\cite{sdedit, stablediffusion}, significantly exacerbates this computational burden.
In this process, a user-provided mask specifies the region to be modified.
As illustrated in Figure~\ref{fig:diffusion_image_editing}(c), while the masked region is synthesized from random noise, the unmasked region is reset at each step by injecting the corresponding timestep noise into the original image to preserve its content.
This process ensures the denoising network focuses on the masked region, effectively preserving the context in surrounding regions.
However, this approach is inherently inefficient.
At every denoising step, the diffusion model processes the entire feature tensor, even though only a small portion undergoes noticeable change.
As a result, a large amount of computation is unnecessarily repeated across regions with negligible changes, leading to significant computational overhead.

To mitigate redundant computation, SIGE~\cite{sige} introduces a caching-based acceleration mechanism.
Specifically, it stores intermediate feature maps from all timesteps generated from the original input image and reuses them for unmasked regions to skip recomputation.
While this design effectively reduces the workload, it requires maintaining large feature tensors derived from a specific reference image.
For Stable Diffusion~\cite{stablediffusion} with a 512$\times$512 image in FP16 precision, the model weights occupy 1.72\,GB, whereas caching feature maps across 50 timesteps demands about 12.31\,GB of memory, 7.16$\times$ larger than the weights themselves.
As a result, its applicability is restricted to scenarios where precomputed features are already available, and the memory footprint grows rapidly with image resolution.
Recent works such as EXION~\cite{exion} have explored dedicated diffusion accelerators to improve general inference efficiency.
However, these accelerators focus on full-image generation and do not account for the redundant computation that arises in masked tasks.
Therefore, a more general and mask-aware solution that operates directly on the input image and its mask is required to achieve both computational efficiency and broad applicability.

\subsection{Quantization}
\begin{figure}[t]
    \centering
    \includegraphics[width=0.9\linewidth]{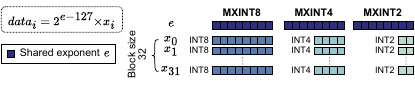}
    \vspace{-0.1in}
    \caption{MXINT8 with proposed extensions MXINT4/2}
    \label{fig:MX}
    \Description{.}
    \vspace{-0.2in}
\end{figure}
Quantization~\cite{han2015deep, jacob2018quantization, nagel2019data, nagel2020up, gholami2022survey, nagel2021white, 10.1145/3579371.3589351} is a widely adopted technique to reduce computational cost by replacing high-precision arithmetic with low-precision numerical representations, such as low-bit integer or floating-point formats.
While simple schemes like tensor-wise scaling are efficient, the strong sensitivity of diffusion models to quantization errors makes coarse schemes insufficient, requiring finer-grained quantization strategies.
In this context, block floating-point (BFP) quantization provides a practical trade-off between numerical precision and hardware efficiency~\cite{msfp, 10.1145/3579371.3589351, mx}.
BFP groups values into blocks, where all values within a block share a single scaling factor, allowing finer scaling granularity compared to tensor- or channel-wise quantization.
The MX format~\cite{mx} is a hardware-friendly BFP implementation where the shared scaling factor is a power-of-two exponent. 
In this work, we extend the MX format by introducing lower-precision variants, MXINT4 and MXINT2, building upon the original MXINT8 as shown in Figure~\ref{fig:MX}.
Each block contains 32 elements that share a shared 8-bit exponent, and the elements are encoded in a two's complement format using INT2, INT4, or INT8 precision.
This format provides an effective balance between hardware efficiency and representational flexibility.
\section{Proposed \HWName: Algorithm}
\label{section3}
\begin{figure}[t]
    \centering
    \includegraphics[width=0.9\linewidth]{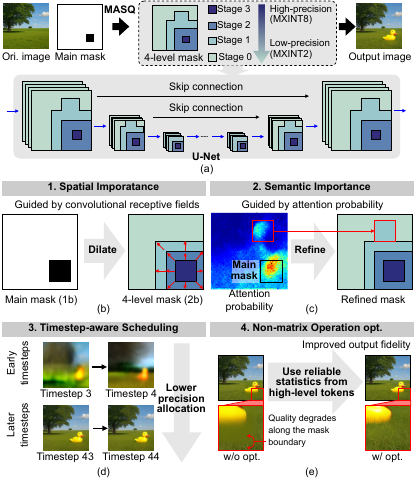}
    \vspace{-0.15in}
    \caption{(a) Overview of the MASQ software method, which allocates multi-precision based on (b) Spatial importance and (c) Semantic importance, along with (d) Timestep-aware precision scheduling and (e) Non-matrix operation optimization}
    \label{fig:algorithm}
    \Description{.}
    \vspace{-0.2in}
\end{figure}
\HWName accelerates masked diffusion through an importance-aware multi-precision algorithm that reduces redundant computation while preserving output quality.
% \HWName accelerates masked diffusion by eliminating redundant computation while preserving output quality.
Figure~\ref{fig:algorithm}(a) illustrates the software algorithm, which assigns multi-level activation precision based on spatial and semantic importance, while keeping all model weights in MXINT8 for numerical stability.
Timestep-aware scheduling enables more aggressive precision reduction, whereas precision-aware normalization and softmax preserve numerical stability during multi-precision execution.

\subsection{Spatial Importance}
Spatial importance stems from the behavior of convolutional receptive fields, which propagate the influence of masked regions into nearby areas.
To define this dependency, we construct a four-stage dilated mask that reflects the spatial extent of these receptive fields.
% Spatial importance is defined through four-stage mask dilation, which reflects the convolutional receptive fields around the masked region.
In diffusion models, 3$\times$3 convolution kernels aggregate features from spatially neighboring tokens, causing the influence of the masked region to extend beyond its explicit boundary.
If masked and unmasked regions are simply assigned high and low precision, respectively, quantization noise from the low-precision region can propagate into adjacent high-precision regions through overlapping receptive fields, degrading reconstruction quality.
This spatial dependency reveals varying levels of importance across regions, where regions closer to the mask boundary require higher precision to preserve visual fidelity.
To capture this property, we employ multi-level quantization using staged mask dilation, which is designed to reflect convolutional receptive fields, assigning higher precision within and around the mask and progressively reducing precision for more distant regions, as illustrated in Figure~\ref{fig:algorithm}(b).

More specifically, we construct a four-stage mask that models the spatial influence range of the user-provided main mask.
Stage 3 corresponds to the main mask itself and represents the core region to be generated.
Stage 2 is obtained by dilating the main mask according to the number of 3$\times$3 convolution layers at the current resolution.
For instance, when two convolution layers are applied, the mask is expanded by two tokens in all directions, including diagonals.
Stage 1 captures inter-resolution influence from the next lower level in the U-Net.
Since downsampling halves the spatial resolution, the dilation distance at this level is doubled to preserve receptive field equivalence.
Stage 0 includes all remaining regions not covered by the previous stages.
By statically analyzing the U-Net architecture, multi-level masks are derived based on the convolutional depth at each resolution.
This spatially aware mask design effectively mitigates error propagation from low-precision computation near masked regions by preserving spatial importance, thereby maintaining overall generation quality.

% \vspace{-0.2in}
\subsection{Semantic Importance}
Beyond spatial masking, we further refine precision allocation by incorporating semantic relevance between tokens.
In diffusion models, tokens in the unmasked region can still influence masked regions through attention, even when spatially distant.
To capture this effect, we quantify semantic importance using attention probabilities, representing the strength of pairwise dependencies computed in self-attention layers.
Specifically, self-attention projects tokens into query and key tensors and computes their dot products, followed by softmax normalization to produce attention probabilities.
As illustrated in Figure~\ref{fig:algorithm}(c), tokens outside the main mask often show high attention probabilities to masked tokens, indicating strong semantic association with masked regions.
Processing such tokens in low precision may cause quantization noise to propagate into the mask through attention layers, degrading the final image quality.

To mitigate this issue, we apply an attention probability–guided refinement that dynamically promotes semantically important tokens to higher precision levels.
During the initial denoising iterations, attention probabilities are computed in the final self-attention layer using MXINT8 and MXINT4 precision to ensure reliable identification of semantically important tokens.
The attention probabilities corresponding to masked tokens are averaged, and if this average exceeds a predefined threshold, tokens initially assigned to Stage 0 are selectively promoted to Stage 1 for higher-precision processing.
By reallocating precision according to semantic relevance, our approach preserves generation fidelity in regions strongly associated with the masked regions through attention.

\begin{figure*}[t]
    \centering
    \includegraphics[width=0.93\linewidth]{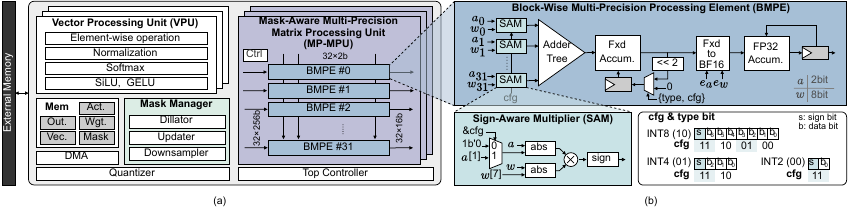}
    \vspace{-0.2in}
    \caption{(a) Overall architecture of \HWName (b) Detailed architecture of BMPE}
    \label{fig:overview of SMAP}
    \Description{.}
    \vspace{-0.1in}
\end{figure*}
\subsection{Timestep-Aware Precision Allocation}
As shown in Figure~\ref{fig:algorithm}(d), diffusion models make large structural updates in early steps and only small adjustments in later steps.
In these later steps, quantization errors have little perceptual impact, allowing lower precision without degrading visual quality.
Motivated by this observation, the precision of unmasked regions is gradually lowered as denoising proceeds, thereby reducing redundant computation.
We empirically determine two downgrade points where the precision configuration transitions from MXINT8/8/4/2 to MXINT8/4/4/2 and finally to MXINT8/4/2/2, reflecting the precision reduction from Stage 3 to Stage 0.
The masked region always remains at MXINT8.  
This simple timestep-aware schedule enables more aggressive precision reduction in later steps, improving efficiency while maintaining visual fidelity.

% \vspace{-0.1in}
\subsection{Precision-Aware Optimization for Non-Matrix Operations}
Quantization errors in diffusion models can also propagate globally through non-matrix operations such as group normalization and softmax.
% Diffusion models also include non-matrix operations such as group normalization and softmax.
Although such non-matrix operations contribute less to overall computation, their global behavior can still propagate quantization errors across tokens, degrading output quality as shown in Figure~\ref{fig:algorithm}(e).
To address this, we incorporate operation-specific optimizations for group normalization and softmax, enabling accurate and stable processing even in multi-precision settings.

For group normalization, the mean and variance are computed across all tokens in a group, and these shared statistics can be easily corrupted by low-precision values, propagating errors to high-precision tokens.
This issue is mitigated by computing normalization statistics only from higher-precision tokens in Stage 2-3 and applying them to all tokens, thereby preserving fidelity in mask regions.
In the softmax operation of self-attention, quantization errors in low-precision keys can distort the attention probabilities of high-precision queries.
To prevent this, the lowest-precision tokens in Stage 0 are excluded from the softmax calculation, effectively assigning them zero probability.
These precision-aware optimizations stabilize normalization and attention under multi-level quantization, preserving overall generation quality.
\vspace{-0.15in}
\label{section4}
\section{Proposed \HWName: Architecture}
\subsection{Overall Architecture}
\HWName is a hardware–software co-designed accelerator that accelerates masked diffusion by efficiently executing the proposed software method, as illustrated in Figure~\ref{fig:overview of SMAP}(a).
\HWName introduces a mask-aware multi-precision execution architecture that dynamically allocates precision according to spatial and semantic importance, enabling efficient processing.
\HWName features a mask-aware multi-precision matrix processing unit (MP-MPU) as its core computational engine, along with a mask manager, a quantizer, and a vector processing unit (VPU).
The MP-MPU consists of 32 block-wise multi-precision processing elements (BMPEs) and executes multi-precision operations guided by stage information from the mask manager.
The quantizer converts activations into a block-wise representation for the MP-MPU, while the VPU handles non-matrix operations, including softmax and normalization.
On-chip memory stores activations, model weights, intermediate features, and mask data, while the top controller manages module execution and the direct memory access (DMA) module transfers data between on-chip and external memory.

\subsection{Multi-Precision Matrix Processing Unit}
The MP-MPU is the core component for multi-precision execution in \HWName, comprising 32 BMPEs that perform block-wise computations at configured precisions.

\begin{figure}[t]
    \centering
    \includegraphics[width=0.9\linewidth]{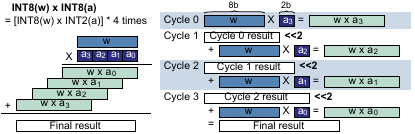}
    % \vspace{-0.1in}
    \caption{Computation flow for multi-precision execution}
    \label{fig:BMPE process}
    \Description{.}
    \vspace{-0.1in}
\end{figure}
\subsubsection{Block-Wise Multi-Precision Processing Element}
Figure~\ref{fig:overview of SMAP}(b) illustrates the architecture of the BMPE, which serves as the fundamental compute unit within the MP-MPU.
Each BMPE processes 32 pairs of 2-bit activations and 8-bit weights in parallel, enabling block-wise computation.
For input activations with 4-bit or 8-bit precision, the BMPE serially processes the data in 2-bit slices.
Each BMPE integrates 32 sign-aware multipliers (SAMs) that perform signed multiplications on activations and weights represented in two's complement form, with partial results accumulated through an adder tree.
To support multi-precision operation, the BMPE receives two control signals: a 2-bit type signal that specifies the precision format --- \textit{00}, \textit{01}, and \textit{10} correspond to MXINT2, MXINT4, and MXINT8 --- and a 2-bit configuration (cfg) signal that indicates which bit slice to process in the current cycle.
Computation begins from the most significant bits (MSBs), with cfg values assigned in descending order of bit significance to control the active slices in each SAM.
Each SAM correctly handles signed multiplication by adjusting the operation based on whether the current 2-bit slice contains a sign bit.
The partial sum from each 2-bit slice is left-shifted according to its bit significance and accumulated in a fixed-point register, as illustrated for an MXINT8 case in Figure~\ref{fig:BMPE process}.
After all slices are processed, the final fixed-point result is first converted to BF16, reflecting the scaling factors of both activation and weight, and then accumulated in FP32 precision to maintain numerical stability.
The accumulated value is finally converted back to BF16 for output.
This bit-serial mechanism allows the BMPE to efficiently support multi-precision computation.
Consequently, an MXINT8 $\times$ MXINT8 operation completes in 4 cycles, while MXINT4 and MXINT2 operations take 2 and 1 cycles, respectively.

\subsubsection{MP-MPU Execution Flow}
The MP-MPU controller determines the data precision for BMPE by combining multi-stage mask information with the current timestep to apply the timestep-aware precision allocation strategy.
% The data precision used in each BMPE is determined by the MP-MPU controller, which receives staged-mask information along with the current timestep to apply the timestep-aware precision allocation strategy.
To support multi-precision load balancing within the MP-MPU, a set of 32 activation pairs is broadcast to all 32 BMPEs.
Each BMPE receives its own set of 32 weights and performs 32 multiplications with the shared activations.
Accordingly, all BMPEs operate under the same precision level and configuration for the current computation cycle.
This execution flow processes all inputs under a unified precision setting, thereby enabling the hardware to natively support region-specific synthesis regardless of the shape or size of the mask.

\subsection{Mask Manager}
\begin{figure}[t]
    \centering
    \includegraphics[width=0.95\linewidth]{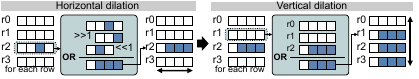}
    % \vspace{-0.1in}
    \caption{Process of mask dilator}
    \label{fig:mask}
    \Description{.}
    \vspace{-0.2in}
\end{figure}
To enable importance-aware precision control, \HWName employs a mask manager that dynamically generates and updates multi-stage masks throughout the denoising process.
The mask manager consists of three submodules: mask dilator, mask updater, and mask downsampler.
It uses a 2-bit encoding scheme for the 4-stage mask, where stage 3, which uses the highest precision, is assigned value \textit{11}, and stages 2, 1, and 0 are assigned \textit{10}, \textit{01}, and \textit{00} respectively.
\subsubsection{Mask Dilator}
The mask dilator generates the initial 4-stage mask by applying spatially guided dilation to the main binary mask.
As illustrated in Figure~\ref{fig:mask}, it sequentially applies horizontal and vertical 1-bit dilation to expand the mask, effectively covering the spatial influence range of convolution layers.
The number of dilation steps is determined statically based on the convolutional structure of the model, pre-profiled before execution.
The dilation module operates on mask tiles of up to 64 $\times$ 64 tokens.
For larger masks, tiling partitions them into 64 $\times$ 64 blocks, enabling reuse of the same hardware logic without modification.
While tiling supports large-scale processing, the dilation logic remains consistent with the spatial influence of convolutional layers.
\subsubsection{Mask Updater}
The mask updater refines precision based on attention probability. 
It receives a binary refinement mask that indicates which tokens exhibit strong semantic associations with the masked tokens, based on a predefined threshold.
For tokens at the lowest precision level, Stage 0, the updater uses a lightweight bitwise OR operation to selectively reassign them to the next precision level, Stage 1.
This process efficiently incorporates semantic importance into the precision assignment with minimal hardware overhead, while leaving Stages 2 and 3 unaffected.
\subsubsection{Mask Downsampler}
To support the U-Net's multi-resolution structure, the mask downsampler generates masks for lower-resolution feature maps.
It reduces the mask resolution by half using a 2$\times$2 sliding window with a stride of 2.
A majority rule is applied to each window, setting the single output bit to one if two or more input bits are one.
This method effectively downsamples the mask while preserving its overall structure across different U-Net resolutions.

\label{section5}
\section{Evaluation}
\subsection{Methodology}
% \begin{table}[htbp]
% \caption{Table Type Styles}
% \begin{center}
% \begin{tabular}{|c|c|c|c|}
% \hline
% \textbf{Table}&\multicolumn{3}{|c|}{\textbf{Table Column Head}} \\
% \cline{2-4} 
% \textbf{Head} & \textbf{\textit{Table column subhead}}& \textbf{\textit{Subhead}}& \textbf{\textit{Subhead}} \\
% \hline
% copy& More table copy$^{\mathrm{a}}$& &  \\
% \hline
% \multicolumn{4}{l}{$^{\mathrm{a}}$Sample of a Table footnote.}
% \end{tabular}
% \label{tab1}
% \end{center}
% \end{table}
\begin{table}[t]
{\footnotesize
\centering
\caption{Model accuracy evaluation}
% \vspace{-0.1in}
\label{table:accuracy}
\begin{tabular}{cc|c|c|c}
\hline
& \textbf{Model} & \multicolumn{2}{c}{\textbf{Stable Diffusion}} & \textbf{SDEdit} \\
\hline
% \multirow{2}{*}{\textbf{Metric}}  & Dataset & \textbf{EditBench} & \textbf{LAION} & \textbf{LSUN Church} \\
% & Edit Ratio & 42.87\% & 2.38\% & 13.12\% \\
{\textbf{Metric}}  & \textbf{Dataset} & \textbf{EditBench} & \textbf{LAION} & \textbf{LSUN Church} \\
\hline
\multirow{2}{*}{IS $\uparrow$} & Original & 5.75 $\pm$ 0.88 & 19.16 $\pm$ 1.51 & 4.02 $\pm$ 0.50 \\
                               & \textbf{Ours}     & 5.94 $\pm$ 0.92 & 19.27 $\pm$ 1.38 & 4.07 $\pm$ 0.44 \\
\hline
\multirow{2}{*}{CLIP $\uparrow$} & Original & 34.19 & 31.18 & - \\
                                 & \textbf{Ours}     & 34.22 & 31.19 & - \\
\hline
\multirow{2}{*}{FID $\downarrow$} & Original & 167.35 & 13.60 & 19.29 \\
                                  & \textbf{Ours}     & 168.73 & 14.07 & 20.91 \\
\hline
\multirow{2}{*}{PSNR $\uparrow$} & Original & 12.81 & 25.15 & 26.28 \\
                                 & \textbf{Ours} & 12.58 & 24.96 & 26.39 \\
% \hline
% \multirow{2}{*}{LPIPS $\downarrow$} & Original & 0.344 & 0.078 & 0.077 \\
%                                     & \textbf{Ours} & 0.346 & 0.080 & 0.080 \\
\hline
\multirow{2}{*}{SSIM $\uparrow$} & Original & 0.563 & 0.820 & 0.900 \\
                                 & \textbf{Ours} & 0.560 & 0.818 & 0.901 \\
\hline
% \vspace{-0.35in}
\end{tabular}
}
\end{table}

To demonstrate the efficiency of \HWName across diverse masked diffusion tasks, we evaluate it on representative cases covering text-guided inpainting with Stable Diffusion v1~\cite{stablediffusion} and stroke-based editing using SDEdit~\cite{sdedit}, which is based on DDPM~\cite{ddpm}, a foundational diffusion model.
The inpainting task is evaluated on two datasets: EditBench~\cite{editbench} (EB), with an average mask ratio of 42.87\%, and an inpainting set from the LAION dataset~\cite{laion}, with a smaller average mask ratio of 2.38\%.
The LAION data was generated following the procedure from SIGE~\cite{sige}.
For the SDEdit task, we use the LSUN Church dataset~\cite{lsun}, with an average mask ratio of 13.12\%, which was prepared for stroke-based editing and publicly released by SIGE~\cite{sige}.
In evaluation, the semantic refinement is applied every five timesteps, and the precision downgrades occur at timesteps 9 and 18 under a 50-timestep schedule.

For hardware evaluation, we implemented \HWName at the RTL level in SystemVerilog and synthesized it using a 14nm process technology, targeting an 800\,MHz operating frequency at 0.8\,V.
We evaluate \HWName against two GPU baselines: a server-class A100~\cite{nvidia_a100} and an edge-class Orin NX~\cite{nvidia_nx}.
We built a cycle-accurate simulator to evaluate the latency and energy consumption of \HWName.
To ensure a fair comparison, we calculate the peak throughput of \HWName under MXINT8$\times$MXINT8 operation to match the FP16 peak performance of A100, 312 TFLOPS, and Orin NX, 3.76 TFLOPS.
\HWName is instantiated in two configurations, an edge version with 102.4 GB/s LPDDR5 memory and a server version with 2 TB/s HBM2E memory.
DRAM energy is estimated based on vendor specifications~\cite{dramvendor, dramvendorlpddr}.

% \vspace{-0.1in}
\subsection{Accuracy}
Table~\ref{table:accuracy} summarizes the accuracy of \HWName under the proposed quantization scheme, compared with the original FP16 model.
Five standard metrics are used: inception score (IS)~\cite{isscore}, CLIP score~\cite{clipscore}, Fréchet inception distance (FID)~\cite{fid}, peak signal-to-noise ratio (PSNR), and structural similarity index measure (SSIM)~\cite{ssim}.
IS and FID evaluate the perceptual quality and distributional similarity, while CLIP assesses semantic alignment with the text prompt.
PSNR quantifies pixel-level fidelity, while SSIM evaluates perceptual similarity at structural levels.
As shown in Table~\ref{table:accuracy}, \HWName achieves accuracy comparable to the FP16 model across all metrics, confirming that \HWName preserves output quality even under aggressive quantization.

\subsection{Performance}
\subsubsection{Speedup and Energy Efficiency}
\begin{figure}[t]
    \centering
    \includegraphics[width=0.9\linewidth]{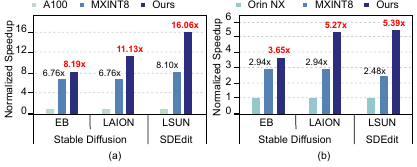}
    \vspace{-0.1in}
    \caption{Speedup comparison with (a) A100 server GPU and (b) Orin NX edge GPU}
    \label{fig:normalized speedup}
    \Description{.}
    % \vspace{-0.1in}
\end{figure}
\begin{figure}[t]
    \centering
    \includegraphics[width=0.9\linewidth]{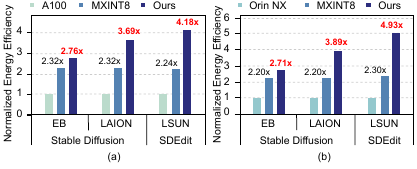}
    \vspace{-0.1in}
    \caption{Energy efficiency comparison with (a) A100 server GPU and (b) Orin NX edge GPU}
    \label{fig:normalized energy efficiency}
    \Description{.}
    % \vspace{-0.3in}
\end{figure}

Figure~\ref{fig:normalized speedup} and Figure~\ref{fig:normalized energy efficiency} present the normalized speedup and energy efficiency of \HWName on both server and edge GPU baselines.
To accurately quantify the benefits of our \HWName algorithm, we first establish a reference point by running the \HWName hardware in a uniformly MXINT8 precision mode.
This approach processes the entire image uniformly, failing to exploit the computational redundancy inherent in masked tasks.

Even without any specialization for masked diffusion, the accelerator's inherent efficiency already delivers substantial gains, achieving up to 8.10$\times$ and 2.94$\times$ speedups and up to 2.32$\times$ and 2.30$\times$ energy efficiency improvements over A100 and Orin NX, respectively.
However, when combined with our scheme specifically optimized to reduce redundant computation in masked diffusion, \HWName achieves greater improvements.
Our mask-aware precision control leads to significant performance gains up to 16.06$\times$ and 5.39$\times$ speedups, along with 4.18$\times$ and 4.93$\times$ energy-efficiency gains over A100 and Orin NX.
On average, this shows a 61.35\% speedup improvement and a 54.88\% energy efficiency gain over the MXINT8 baseline on the server configuration.
On the edge configuration, \HWName achieves an average speedup improvement of 73.58\% and an average energy efficiency gain of 71.45\% over the MXINT8 baseline.

Notably, GPUs and the baseline MXINT8 configuration remain static regardless of mask size.
In contrast, \HWName leverages mask information to dynamically adjust its precision based on mask pattern, enabling significantly greater performance gains.
Moreover, the robust improvements on both server and edge GPUs demonstrate that \HWName remains effective across diverse inference environments.

\subsubsection{Latency}
\begin{figure}[t]
    \centering
    \includegraphics[width=0.9\linewidth]{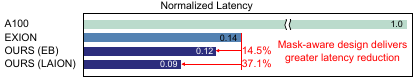}
    \vspace{-0.1in}
    \caption{Latency comparison with A100 and EXION}
    \label{fig:latency}
    \Description{.}
    % \vspace{-0.1in}
\end{figure}

Figure~\ref{fig:latency} presents the latency comparison of \HWName against A100 and a prior SOTA diffusion accelerator, EXION~\cite{exion}, which is designed for general diffusion models rather than masked diffusion.
The comparison is based on Stable Diffusion results, where the value for EXION is taken from the original paper~\cite{exion}.
Compared to EXION, which achieves an 85.7\% latency reduction, \HWName further reduces latency by 14.5\% and 37.1\% on the EB and LAION datasets, respectively, demonstrating superior efficiency in handling masked diffusion workloads.
\HWName demonstrates better performance on the LAION dataset due to its smaller editing regions, which allow more aggressive low-precision quantization.
These results demonstrate the advantage of the mask-aware multi-precision execution of \HWName in accelerating masked diffusion workloads beyond conventional diffusion accelerators.

\subsection{Area and Power Breakdown}
\begin{table}[t]
{\footnotesize
\centering
\caption{Breakdown of area and power}
\vspace{-0.1in}
\label{table:area_power_breakdown}
\begin{tabular}{lcc}
\hline
\textbf{Component} & \textbf{Area (mm$^2$)} & \textbf{Power (W)} \\ \hline
MP-MPU & 1.91 & 2.29 \\
VPU & 1.27 & 2.14 \\
Mask Manager \& Quantizer & 0.07 & 0.12 \\
On-chip Memories & 2.28 & 0.78 \\
MISC & 0.62 & 0.35 \\ \hline
\textbf{Total} & \textbf{6.15} & \textbf{5.68} \\ \hline
% \vspace{-0.3in}
\end{tabular}}
\end{table}
% \begin{table}[t]
% \centering
% \caption{Breakdown of area and power}
% \label{table:area_power_breakdown}
% \begin{tabular}{lcc}
% \hline
% \textbf{Component} & \textbf{Area (mm$^2$)} & \textbf{Power (W)} \\ \hline
% MP-MPU & 51.81 & 63.40 \\
% Mask Manager & 0.27 & 0.41 \\
% Vector Processor Unit & 30.36 & 51.45 \\
% Quantizer & 1.40 & 2.47 \\
% Memory & 9.78 & 3.24 \\
% MISC & 4.69 & 4.88 \\ \hline
% \textbf{Total} & \textbf{98.31} & \textbf{125.87} \\ \hline
% \end{tabular}
% \end{table}

Table~\ref{table:area_power_breakdown} details the area and power breakdown of \HWName, synthesized at 800\,MHz and 0.8\,V.
In the synthesized configuration with 32 MP-MPUs and a 2\,MiB on-chip memory, each MP-MPU consisting of 32 BMPEs, it has an area of 6.15\,mm$^2$ and a power consumption of 5.68\,W.
The MP-MPU constitutes the dominant portion of hardware cost, while the mask manager and quantizer add only 1.13\% area and 2.11\% power, although they are essential components of the design.
This highlights that our approach delivers substantial performance gains while imposing minimal additional resource overhead.
\label{section6}
\section{Conclusion}
This paper presents \HWName, a hardware-software co-designed accelerator for efficient masked diffusion.
At the software level, \HWName integrates a stage-wise multi-precision quantization that dynamically allocates precision based on spatial and semantic importance derived from the mask, along with timestep-aware scheduling and non-matrix optimization to eliminate redundant computation.
At the hardware level, \HWName incorporates a flexible multi-precision compute engine and specialized mask management logic, enabling efficient execution of this scheme with minimal overhead.
Compared with the NVIDIA A100 and Jetson Orin NX, \HWName achieves up to 16.06$\times$ and 5.39$\times$ speedup and up to 4.18$\times$ and 4.93$\times$ energy efficiency improvement, respectively, without quality degradation.
\section*{Acknowledgment}
This work was supported by the Institute of Information \& communications Technology Planning \& Evaluation (IITP) through the Graduate School of AI Semiconductor (IITP-2026-RS-2023-00256472), and the IITP-ITRC program (IITP-2026-RS-2020-II201847), all funded by the Korea government (MSIT). The EDA tool was supported by the IC Design Education Center(IDEC), Korea.

\newpage
\bibliographystyle{ACM-Reference-Format} 
\bibliography{Thesis/99_reference}

\end{document}